\pdfoutput=1
\RequirePackage{ifpdf}
\ifpdf 
\documentclass[pdftex]{sigma}
\else
\documentclass{sigma}
\fi

\begin{document}

\newcommand\su{\mathfrak{su}}
\newcommand\osp{\mathfrak{osp}}
\newcommand{\myatop}[2]{\genfrac{}{}{0pt}{}{#1}{#2}}


\renewcommand{\PaperNumber}{025}

\FirstPageHeading

\ShortArticleName{Deformed $\su(1,1)$ Algebra as a Model for Quantum Oscillators}

\ArticleName{Deformed $\boldsymbol{\su(1,1)}$ Algebra\\ as a Model for Quantum Oscillators}

\Author{Elchin I. JAFAROV~$^{\dag\ddag}$,
Neli I. STOILOVA~$^\S$ and Joris VAN DER JEUGT~$^\dag$}

\AuthorNameForHeading{E.I.~Jafarov, N.I.~Stoilova and J.~Van der Jeugt}

\Address{$^\dag$~Department of Applied Mathematics and Computer Science, Ghent University,\\
\hphantom{$^\dag$}~Krijgslaan 281-S9, B-9000 Gent, Belgium}
\EmailD{\href{mailto:Joris.VanderJeugt@UGent.be}{Joris.VanderJeugt@UGent.be}}

\Address{$^\ddag$~Institute of Physics, Azerbaijan National Academy of Sciences,\\
\hphantom{$^\ddag$}~Javid Av. 33, AZ-1143 Baku, Azerbaijan}
\EmailD{\href{mailto:ejafarov@physics.ab.az}{ejafarov@physics.ab.az}}

\Address{$^\S$~Institute for Nuclear Research and Nuclear Energy,\\
\hphantom{$^\S$}~Boul.\ Tsarigradsko Chaussee 72, 1784 Sofia, Bulgaria}
\EmailD{\href{mailto:stoilova@inrne.bas.bg}{stoilova@inrne.bas.bg}}

\ArticleDates{Received February 17, 2012, in f\/inal form May 08, 2012; Published online May 11, 2012}

\Abstract{The Lie algebra $\mathfrak{su}(1,1)$ can be deformed by a ref\/lection operator, in such a way that
the positive discrete series representations of~$\mathfrak{su}(1,1)$ can be extended to representations
of this deformed algebra $\mathfrak{su}(1,1)_\gamma$.
Just as the positive discrete series representations of~$\mathfrak{su}(1,1)$ can be used to model
a quantum oscillator with Meixner--Pollaczek polynomials as wave functions,
the corresponding representations of $\mathfrak{su}(1,1)_\gamma$ can be utilized to construct
models of a quantum oscillator.
In this case, the wave functions are expressed in terms of continuous dual Hahn polynomials.
We study some properties of these wave functions, and illustrate some features in plots.
We also discuss some interesting limits and special cases of the obtained oscillator models.}


\Classification{81R05; 81Q65; 33C45}

\section{Introduction}

The $\su(1,1)$-model of a quantum oscillator~\cite{Klimyk2006} is a model that obeys the dynamics
of a harmonic oscillator, but with the position and momentum operators and the Hamiltonian being elements of
the Lie algebra $\su(1,1)$ instead of the Heisenberg algebra.

There are many algebraic constructions to model a quantum oscillator.
The dif\/f\/iculty for such models is often to determine the spectra of observables and an explicit form of
their eigenfunctions.
Only for some models, one can develop such a complete theory.
One of these models is the $q$-oscillator, a $q$-deformation of the standard quantum
oscillator~\cite{Biedenharn1989,Klimyk2005,Macfarlane1989,Sun1989}.
The $q$-oscillator has many interesting properties, both from the mathematics and physics point of view.
But it also has some drawbacks, in particular the Newton--Lie (or Hamilton--Lie) equations are not satisf\/ied.

Following this, new oscillator models were developed such that the same dynamics as in the
classical or quantum case is satisf\/ied, and in such a way that the
operators correspon\-ding to position, momentum and Hamiltonian are elements of some algebra dif\/ferent
from the traditional Heisenberg (or oscillator) Lie algebra.
In the one-dimensional case, there are three (essentially self-adjoint) operators:
a~position operator~$\hat q$, its corresponding momentum opera\-tor~$\hat p$ and
a~(pseudo-)Hamiltonian $\hat H$ which is the generator of time evolution.
These operators should satisfy the Hamilton--Lie equations (or the compatibility of Hamilton's equations with the Heisenberg
equations):
\begin{equation}
[\hat H, \hat q] = -i \hat p, \qquad [\hat H,\hat p] = i \hat q,
\label{Hqp}
\end{equation}
in units with mass and frequency both equal to~1, and $\hbar=1$.
Contrary to the canonical case, the commutator $[\hat q, \hat p]=i$ is not required.
Apart from~\eqref{Hqp} and the self-adjointness, it is then common to require the following conditions~\cite{Atak2001}:
\begin{itemize}\itemsep=0pt
\item all operators $\hat q$, $\hat p$, $\hat H$ belong to some (Lie) algebra (or superalgebra) $\cal A$;
\item the spectrum of $\hat H$ in (unitary) representations of $\cal A$ is equidistant.
\end{itemize}
A very interesting model occurs for ${\cal A}= \su(2)$ (or its enveloping algebra)~\cite{Atak2001,Atak2001b,Atak2005}.
In that case, the relevant representations are the well known~$\su(2)$ representations
labeled by an integer or half-integer~$j$.
Since these representations are f\/inite-dimensional, one is dealing with ``f\/inite oscillator models'', of
potential use in optical image processing~\cite{Atak2005}.
Up to a constant, the Hamiltonian $\hat H$ is the diagonal $\su(2)$ operator with a linear spectrum $n+\frac12$ ($n=0,1,\ldots,2j$).
Also $\hat q$ and $\hat p$ have a f\/inite spectrum, given by $\{-j,-j+1,\ldots,+j\}$~\cite{Atak2001}.
The (discrete) position wave functions have been constructed, and are given by Krawtchouk functions (normalized
symmetric Krawtchouk polynomials)~\cite{Atak2001}, tending to the
canonical wave functions in terms of Hermite polynomials when $j\rightarrow \infty$.
In the terminology of quantum theory of angular momentum, these discrete position wave functions
are just Wigner D-functions~\cite{Atak-Suslov}, and their relation to Krawtchouk polynomials
was f\/irst given by Koornwinder~\cite{Koornwinder1982}.

In two previous papers~\cite{JSV2011,JSV2011b}, the $\su(2)$ model for the f\/inite one-dimensional harmonic oscillator
was extended. The underlying algebra is a deformation of $\su(2)$ with
an extra ref\/lection (or parity) operator and an additional parameter $\alpha$ $({>}{-}1)$.
In the even-dimensional representations~\cite{JSV2011} ($j$ half-integer),
the spectrum of the position operator is of the form
\begin{equation*}
-\alpha-j-\frac12, -\alpha-j+\frac12, \ldots, -\alpha-1; \alpha+1,\alpha+2, \ldots,\alpha+j+\frac12,
\end{equation*}
and the position wave functions could be constructed in terms of normalized Hahn (or dual Hahn) polynomials
(with parameters $(\alpha,\alpha+1)$ or $(\alpha+1,\alpha)$).
In the odd-dimensional representations~\cite{JSV2011b} ($j$ integer), the spectrum of the position operator is
\[
0, \quad \pm \sqrt{k(2\alpha+k+1)}, \qquad k=1,\ldots,j.
\]
The position (and momentum) wave functions are again Hahn
polynomials (in this case with parameters $(\alpha,\alpha)$ or $(\alpha+1,\alpha+1)$).

Models of quantum oscillators with continuous spectra of position and momentum operators were
constructed based on the positive discrete series representations of $\su(1,1)$~\cite{Klimyk2006}.
In such a~representation, labeled by a positive number $a>0$, the spectrum of the position operator
is~${\mathbb R}$. The position wave function, when the oscillator is in the $n$th eigenstate of the Hamiltonian,
is given by
\begin{equation*}
\phi_n^{(a)} (x) = \frac{2^a \sqrt{n!}}{\sqrt{2\pi\, \Gamma(n+2a)}} |\Gamma(a+ix)|\, P_n^{(a)}(x;\pi/2),
\end{equation*}
where $P_n^{(\lambda)}(x;\phi)$ is the Meixner--Pollaczek polynomial~\cite{Koekoek}.
Many interesting properties of these $\su(1,1)$ oscillators were described by Klimyk~\cite{Klimyk2006}.

One type of deformation of this $\su(1,1)$ model was of\/fered by its $q$-deformation.
The $\su_q(1,1)$ model was investigated in~\cite{Atak2006}.
The position and momentum operators have spectra covered by a~f\/inite interval of the real line, which depends on
the value of $q$, and the wavefunctions are given in terms of $q$-Meixner--Pollaczek polynomials.

In the current paper, we consider an extension of the Lie algebra $\su(1,1)$ by a parity or ref\/lection operator $R$.
In this extension or deformation, the common $\su(1,1)$ commutator $[J_+,J_-]=-2J_0$ is replaced by
$[J_+,J_-]=-2J_0-\gamma R$, where $\gamma$ is a (real) deformation parameter.
The positive discrete series representations of $\su(1,1)$, labeled by a positive number $a$, can
be extended to representations of the deformed algebra $\su(1,1)_\gamma$, provided $\gamma$ can be written
in the form $\gamma=(2a-1)(2c-1)$ for some positive $c$-value (sometimes $c$ rather than $\gamma$ will be referred to
as the deformation parameter).
Section~\ref{section2} recalls some known formulas for the $\su(1,1)$ case, and in Section~\ref{section3} the algebra $\su(1,1)_\gamma$
and its representations are given.
The core of the paper comes in Section~\ref{section4}, where models for a quantum oscillator are built using $\su(1,1)_\gamma$
representations.
Just as for the canonical oscillator, the Hamiltonian has a discrete but inf\/inite equidistant spectrum in these models,
and the position operator has spectrum~${\mathbb R}$.
We have managed to obtain explicit expressions for the orthonormal wave functions~$\psi_n^{(a,c)}(x)$,
where~$a$ is the representation label and~$c$ the deformation label.
These wave functions involve the class of so-called continuous dual Hahn polynomials~\cite{Koekoek, Ismail}.
Their properties and shapes are studied in Section~\ref{section4}.
In Section~\ref{section5} we consider some limits and special cases. For $c=1/2$ (or $\gamma=0$), our functions reduce to
those studied by Klimyk~\cite{Klimyk2006}. Another interesting case is when~$c$ tends to~$+\infty$: then
the model reduces to that of the paraboson oscillator. So for $a=1/2$ and $c\rightarrow +\infty$ the model
coincides with the canonical oscillator.
In Section~\ref{section6}, we give a dif\/ferential-ref\/lection operator realization of the deformed algebra $\su(1,1)_\gamma$,
and in Section~\ref{section7} we discuss the possibility of considering further parameters in the deformation.
Our paper closes with some remarks in a~concluding section.

\section[The algebra $\su(1,1)$, positive discrete series representations and oscillator models]{The algebra $\boldsymbol{\su(1,1)}$, positive discrete series representations\\ and oscillator models}\label{section2}

The Lie algebra $\su(1,1)$~\cite{Klimyk2006} can be def\/ined by its basis elements
$J_0$, $J_+$, $J_-$ with commutator relations
\begin{equation*}
[J_0,J_\pm]=\pm J_\pm, \qquad [J_+,J_-]=-2J_0.
\end{equation*}
The positive discrete series representations of $\su(1,1)$ are labeled~\cite{Bargmann, Klimyk2006} by a positive
real number $a>0$ (the Bargmann index), and are inf\/inite-dimensional.
The action of the $\su(1,1)$ generators on a set of basis vectors $|a,n\rangle$ (with $n=0,1,2,\ldots$) is given by
\begin{gather}
  J_0 |a,n\rangle = (n+a)\,|a,n\rangle,\nonumber\\
  J_+ |a,n\rangle = \sqrt{(n+1)(n+2a)}\,|a,n+1\rangle, \label{su11-act}\\
  J_- |a,n\rangle = \sqrt{n(n+2a-1)}\, |a,n-1\rangle. \nonumber
\end{gather}
This action satisf\/ies $J_0^\dagger=J_0$, $J_\pm^\dagger=J_\mp$.
The representation space ${\cal H}_a$ is a Hilbert space with orthonormal basis $|a,n\rangle$ (with $n=0,1,2,\ldots$).

In the $\su(1,1)$ oscillator model, the position, momentum and Hamiltonian are chosen as follows:
\begin{equation}
\hat q = \frac12 (J_++J_-), \qquad
\hat p = \frac{i}{2}(J_+-J_-), \qquad
\hat H = J_0 . \label{su11-qpH}
\end{equation}
These operators satisfy~\eqref{Hqp}. The spectrum of $\hat H$ is thus $(n+a)$ ($n=0,1,2,\ldots$), an equidistant
and inf\/inite spectrum. One can just as well choose the representation-dependent operator $J_0-a+1/2$ for $\hat H$
in order to get the spectrum of the standard quantum oscillator, but in fact this shift is not so relevant.
A more interesting aspect is the determination of the spectrum of the position operator $\hat q$ and its
eigenvectors.
In fact, the matrix elements for $\hat q$ can be deduced from the explicit computations of Basu and
Wolf~\cite{Basu}, and were (for the current representations) identif\/ied with Meixner--Pollaczek polynomials
by Koornwinder~\cite{Koornwinder1988}.
Let us brief\/ly describe the method followed by Klimyk~\cite{Klimyk2006}.
Denoting a formal eigenvector of $\hat q$, for the eigenvalue $x$, by
\[
v(x) = \sum_{n=0}^\infty A_n(x)\, |a,n\rangle,
\]
the equation ${\hat q}  v(x) = x   v(x)$ leads by means of \eqref{su11-act} and~\eqref{su11-qpH} to
\begin{equation*}
2x  A_n(x) = \sqrt{(n+1)(n+2a)} A_{n+1}(x) + \sqrt{n(n+2a-1)} A_{n-1}(x).
\end{equation*}
It was shown in~\cite{Klimyk2006,Koelink1998} that this recurrence relation is satisf\/ied by normalized Meixner--Pollaczek
polynomials $P_n^{(a)}(x;\pi/2)$, i.e.
\begin{equation}
A_n(x) = \phi_n^{(a)} (x) = \frac{2^a \sqrt{n!}}{\sqrt{2\pi  \Gamma(n+2a)}} |\Gamma(a+ix)|  P_n^{(a)}(x;\pi/2),
\label{phi-n}
\end{equation}
where these Meixner--Pollaczek polynomials are given in terms of the hypergeometric function as
\begin{equation}
P_n^{(a)}(x;\pi/2) = \frac{(2a)_n}{n!} i^n \, {}_2F_1 \left( \myatop{-n,a+ix}{2a} ; 2 \right).
\label{defP}
\end{equation}
We have used here the common notation for Pochhammer symbols~\cite{Andrews,Bailey,Slater}
$(a)_k=a(a+1)\cdots(a+k-1)$ for $k=1,2,\ldots$ and $(a)_0=1$.
These (real) functions are orthonormal,
\[
\int_{-\infty}^{+\infty} \phi_n^{(a)} (x) \phi_m^{(a)} (x) = \delta_{nm},
\]
with support ${\mathbb R}$. Hence the spectrum of $\hat q$ is ${\mathbb R}$~\cite{Berezanskii, Klimyk2006, Regniers2010}, and
the functions~\eqref{phi-n} have an interpretation as position wave functions~\cite{Regniers2010}.

\section[The deformed algebra $\su(1,1)_\gamma$ and its representations]{The deformed algebra $\boldsymbol{\su(1,1)_\gamma}$ and its representations}\label{section3}

The extension and deformation of the Heisenberg algebra with a ref\/lection operator was performed in~\cite{Ply}.
In that case, the representations of the deformed algebra correspond to representation of the Lie superalgebra $\osp(1|2)$.
Recently, there has been further interest in extending algebras by ref\/lection operators~\cite{Vinet1},
or even more basically in extending dif\/ferential operators by ref\/lection operators~\cite{Post, Vinet2}.
It is in this context that we present an extension and deformation of $\su(1,1)$.

The universal enveloping algebra of $\su(1,1)$ can be extended by a parity (or ref\/lection) operator $R$,
with action $R |a,n\rangle = (-1)^{n}\;|a,n\rangle$ in the representation ${\cal H}_a$. This means
that $R$ commutes with $J_0$, anticommutes with $J_+$ and $J_-$, and $R^2=1$.
This extended algebra can be deformed by a parameter $\gamma$, leading to the def\/inition of $\su(1,1)_\gamma$.

\begin{definition}\label{definition1}
Let $\gamma$ be a parameter. The algebra $\su(1,1)_\gamma$ is a unital algebra with basis ele\-ments~$J_0$, $J_+$, $J_-$ and $R$ subject to the following relations:
\begin{itemize}\itemsep=0pt
\item $R$ is a parity operator satisfying $R^2=1$ and
\begin{equation}
[R,J_0]=RJ_0-J_0R=0, \qquad \{R,J_\pm\}=RJ_\pm + J_\pm R= 0.
\label{R}
\end{equation}
\item The $\su(1,1)$ relations are deformed as follows:
\begin{gather}
  [J_0, J_\pm] = \pm J_\pm,  \label{J0J+} \\
 [J_+, J_-] = -2 J_0 -\gamma R.
\label{J+J-}
\end{gather}
\end{itemize}
\end{definition}

Clearly, for $\gamma=0$ this is just $\su(1,1)$ extended by a parity operator, and the positive discrete
series representations ${\cal H}_a$ are the same as~\eqref{su11-act} with extra action $R \,|a,n\rangle = (-1)^{n}\,|a,n\rangle$.
The interesting property is that also for $\gamma\ne 0$, the representation ${\cal H}_a$ ($a>0$) can be deformed in
such a way that it becomes a representation for~$\su(1,1)_\gamma$, provided some conditions are satisf\/ied for $a$.
This is described in the following proposition. From now on we shall assume that~$\gamma$ is a~given nonzero real number.

\begin{proposition}\label{proposition1}
Let $a$ be a positive real number with $a\ne 1/2$ and consider the space ${\cal H}_a$ with basis vectors
$|a,n\rangle$ $(n=0,1,2,\ldots)$. Assume that $\gamma$ can be written as
\begin{equation}
\gamma=(2a-1)(2c-1)\qquad \hbox{with} \quad c>0.
\label{ac}
\end{equation}
Then the following action turns ${\cal H}_a$ into
an irreducible representation space of $\su(1,1)_\gamma$
\begin{gather}
 R |a,n\rangle = (-1)^{n}\,|a,n\rangle,\label{act-R}\\
  J_0 |a,n\rangle = \left(n+a+c-\frac12\right) |a,n\rangle,\label{act-J0}\\
  J_+ |a,n\rangle =
  \begin{cases}
 \sqrt{(n+2a)(n+2c)}\,|a,n+1\rangle, & \text{if $n$ is even,}\\
 \sqrt{(n+1)(n+2a+2c-1)}\,|a,n+1\rangle, & \text{if $n$ is odd,}
 \end{cases} \label{act-J+}\\
  J_- |a,n\rangle =
  \begin{cases}
 \sqrt{n(n +2a+2c-2)}\,|a,n-1\rangle, & \text{if $n$ is even,}\\
 \sqrt{(n+2a-1)(n+2c-1)}\,|a,n-1\rangle, & \text{if $n$ is odd.}
 \end{cases} \label{act-J-}
\end{gather}
\end{proposition}

So, for a given $\gamma\ne 0$ (i.e.\ for a f\/ixed algebra $\su(1,1)_\gamma$), not all positive $a$-values are
allowed as representation parameter. From~\eqref{ac}, it follows that:
\begin{itemize}\itemsep=0pt
\item if $\gamma<0$ then $a \in ]0,\frac12 [ \cup ]\frac{1-\gamma}{2}, +\infty[$,
\item if $0<\gamma<1$ then $a \in ]0,\frac{1-\gamma}{2} [ \cup ]\frac12, +\infty[$,
\item if $\gamma\geq 1$ then $a \in ]\frac12 , +\infty[$.
\end{itemize}

The proof of Proposition~\ref{proposition1} is essentially by direct computation: it is a simple task to
verify that the actions~\eqref{act-R}--\eqref{act-J-} do indeed satisfy the relations~\eqref{R}--\eqref{J+J-}.
The conditions $a>0$ and $c>0$ ensure that the factors under the square roots are positive.
The irreducibility follows from the fact that $(J_+)^m \,|a,n\rangle$ is nonzero and proportional to $|a,n+m\rangle$
for $n,m=0,1,2,\ldots$, and that $(J_-)^m \,|a,n+m\rangle$ is nonzero and proportional to $|a,n\rangle$.

Note also that the representation given in this proposition is unitary under the star conditions
$R^\dagger=R$, $J_0^\dagger=J_0$, $J_\pm^\dagger=J_\mp$.

\section[A one-dimensional oscillator model based on $\su(1,1)_\gamma$]{A one-dimensional oscillator model based on $\boldsymbol{\su(1,1)_\gamma}$}\label{section4}

Following the choice in~\cite{Klimyk2006} (as in Section~\ref{section2})
let us choose the position, momentum and Hamiltonian operators as follows:
\begin{equation}
\hat q = \frac12 (J_++J_-), \qquad
\hat p = \frac{i}{2}(J_+-J_-), \qquad
\hat H = J_0 -\left(c-\frac12\right) .
\label{qpH}
\end{equation}
with $c$ determined by~\eqref{ac}.
The choice of subtracting a constant $c-\frac12$ is just for convenience, and in fact this shift
in the spectrum is not so relevant.
The operators~\eqref{qpH} satisfy~\eqref{Hqp}.
In the representation space ${\cal H}_a$, $\hat H \,|a,n\rangle = (n+a)\,|a,n\rangle$,
therefore the spectrum of $\hat H$ is linear and given~by
\begin{equation*}
n+a, \qquad  n=0,1,2,\ldots.
\end{equation*}

Just as for the undeformed $\su(1,1)$ models, an interesting question is the
determination of the spectrum of the position operator $\hat q$ (resp.\ momentum operator $\hat p$) and its
eigenvectors.
Writing the action of $\hat q$ on even and odd states separately, one has from~\eqref{act-J+}--\eqref{act-J-}:
\begin{gather}
  {\hat q}|a,2n\rangle  = \sqrt{n(n+a+c-1)} \,|a,2n-1\rangle + \sqrt{(n+a)(n+c)} \,|a,2n+1\rangle, \nonumber\\
  {\hat q}|a,2n+1\rangle  = \sqrt{(n+a)(n+c)} \,|a,2n\rangle + \sqrt{(n+1)(n+a+c)} \,|a,2n+2\rangle.
\label{act-q}
\end{gather}
If we denote the formal eigenvector of $\hat q$, for the eigenvalue $x$, again by
\begin{equation}
v(x) = \sum_{n=0}^\infty A_n(x)\, |a,n\rangle,
\label{v(x)}
\end{equation}
the equation ${\hat q}  v(x) = x  v(x)$ leads due to~\eqref{act-q} to
\begin{gather}
  x A_{2n}(x) = \sqrt{(n+a)(n+c)} A_{2n+1}(x) + \sqrt{n(n+a+c-1)} A_{2n-1}(x),\label{A1}\\
  x  A_{2n+1}(x) = \sqrt{(n+1)(n+a+c)} A_{2n+2}(x) + \sqrt{(n+a)(n+c)} A_{2n}(x). \label{A2}
\end{gather}
The task is now to solve~\eqref{A1}--\eqref{A2}.
The solution can again be found in the context of orthogonal polynomials.
However, \eqref{A1}--\eqref{A2} is not simply the recurrence relation of a known (normalized) set of orthogonal
polynomials. Instead, we will have to combine two sets of such orthogonal polynomials.

For this purpose, let us recall the continuous dual Hahn polynomials $S_n(x^2;a;b;c)$ in the variable $x^2$, with
at least two positive parameters $a$, $b$ and $c$~\cite{Andrews,Ismail, Koekoek}.
These polynomials of degree $n$ ($n=0,1,2,\ldots$) in the variable $x^2$
are def\/ined by~\cite{Ismail, Koekoek}:
\begin{equation}
\frac{S_n(x^2;a,b,c)}{(a+b)_n(a+c)_n} = {}_3F_2 \left( \myatop{-n,a+ix,a-ix}{a+b,a+c} ; 1 \right),
\label{defS}
\end{equation}
in terms of the generalized hypergeometric series $_3F_2$ of unit argument~\cite{Andrews,Bailey,Slater}.
For $a,c>0$ and $b\geq 0$, the orthogonality relation of these polynomials reads~\cite{Groenevelt,Ismail, Koekoek}:
\begin{gather}
 \frac{1}{2\pi} \int_0^{+\infty} \left| \frac{\Gamma(a+ix)\Gamma(b+ix)\Gamma(c+ix)}{\Gamma(2ix)} \right|^2
S_m(x;a,b,c) S_n(x;a,b,c)\, dx \nonumber\\
\qquad{} = \Gamma(n+a+b)\Gamma(n+a+c)\Gamma(n+b+c) n!  \delta_{mn}.
\label{orth-S}
\end{gather}
Like all orthogonal polynomials, the continuous dual Hahn polynomials satisfy a 3-term recurrence relation.
However, it is not this relation that can be used here.
Instead, we shall need to make use of dif\/ference relations, given in the following proposition.
Note that these dif\/ference relations appeared already in~\cite[equation~(2.13)]{Groenevelt};
we give a more general proof here.
\begin{proposition}\label{proposition2}
Continuous dual Hahn polynomials satisfy the following difference relations:
\begin{gather}
  \left(x^2+b^2\right) S_n\left(x^2;a,b+1,c\right) = (n+a+b)(n+b+c) S_n\left(x^2;a,b,c\right)-S_{n+1}\left(x^2;a,b,c\right),\!\!\!\!  \label{S1}\\
  S_n\left(x^2;a,b,c\right) = S_n\left(x^2;a,b+1,c\right)-n(n+a+c-1)S_{n-1}\left(x^2;a,b+1,c\right). \label{S2}
\end{gather}
\end{proposition}

\begin{proof}
Let us start with the following identity:
\[
{}_3F_2 \left( \myatop{-n,A,B}{C,D} ; z \right) - {}_3F_2 \left( \myatop{-n-1,A,B}{C,D} ; z \right)=
\frac{zAB}{CD} \,{}_3F_2 \left( \myatop{-n,A+1,B+1}{C+1,D+1} ; z \right).
\]
To see that this identity holds, compare coef\/f\/icients of $z^k$ in left and right hand side.
Putting $A=a+ix$, $B=a-ix$, $C=a+b$, $D=a+c$, $z=1$ and following~\eqref{defS}, this can be written as
\[
\left(a^2+x^2\right) S_n\left(x^2;a+1,b,c\right) = (n+a+b)(n+a+c) S_n\left(x^2;a,b,c\right)-S_{n+1}\left(x^2;a,b,c\right).
\]
Since the continuous dual Hahn polynomials are symmetric in $(a,b,c)$, \eqref{S1} follows
by permu\-ting~$a$ and~$b$.
To prove~\eqref{S2}, one can start from the following contiguous relation between ${}_3F_2$'s:
\begin{equation}
(n+C)\,{}_3F_2 \left( \myatop{-n,A,B}{C+1,D} ; z \right) - n \,{}_3F_2 \left( \myatop{-n+1,A,B}{C+1,D} ; z \right)=
 C \,{}_3F_2 \left( \myatop{-n,A,B}{C,D} ; z \right). \label{contiguous}
\end{equation}
To verify that this relation is correct, one simply compares powers of $z^k$ in the left and right hand side.
Finally, putting $A=a+ix$, $B=a-ix$, $C=a+b$, $D=a+c$ and $z=1$ in~\eqref{contiguous} yields~\eqref{S2}.
\end{proof}

In particular, it follows from the previous proposition that
\begin{gather}
  x^2 S_n\left(x^2;a,1,c\right) = (n+a)(n+c) S_n\left(x^2;a,0,c\right)-S_{n+1}\left(x^2;a,0,c\right),  \label{S10}\\
  S_n(x^2;a,0,c) = S_n\left(x^2;a,1,c\right)-n(n+a+c-1)S_{n-1}\left(x^2;a,1,c\right). \label{S20}
\end{gather}
Comparing with the relations~\eqref{A1}--\eqref{A2}, we can propose:
\begin{gather*}
  A_{2n}(x)= \frac{(-1)^n S_n(x^2;a,0,c)}{\sqrt{\Gamma(n+a)\Gamma(n+c)\Gamma(n+a+c) n!}},  \\ 
  A_{2n+1}(x)= \frac{(-1)^n   x  S_n(x^2;a,1,c)}{\sqrt{\Gamma(n+a+1)\Gamma(n+c+1)\Gamma(n+a+c) n!}}. 
\end{gather*}
It is now rather trivial to see that \eqref{A1} follows from~\eqref{S20}, and that \eqref{A2} follows from~\eqref{S10}.
So we have found a solution for the recurrence relations~\eqref{A1}--\eqref{A2}.
Note that the solution is in terms of {\em two} sets of orthogonal polynomials: continuous dual Hahn polynomials
with parameters $(a,0,c)$ for the even degree polynomials, and with parameters $(a,1,c)$ for the odd
degree polynomials.

The expression for $A_n(x)$ is a real polynomial of degree~$n$ in $x$. For the appropriate function
\begin{equation*}
w(x)=\frac{1}{4\pi}\left| \frac{\Gamma(a+ix)\Gamma(ix)\Gamma(c+ix)}{\Gamma(2ix)} \right|^2  
\end{equation*}
we have
\begin{equation}
\int_{-\infty}^{+\infty} w(x) A_m(x) A_n(x) dx = \delta_{mn}.  \label{A-orth}
\end{equation}
Indeed, for $m$ and $n$ even, one writes $\int_{-\infty}^{+\infty} = 2 \int_{0}^{+\infty}$, and then the result
follows directly from~\eqref{orth-S}. For $m$ and $n$ odd, one uses again $\int_{-\infty}^{+\infty} = 2 \int_{0}^{+\infty}$.
The extra appearance of $x^2$ as a factor is taken care of by
$|\Gamma(ix)|^2 x^2 = | ix\Gamma(ix) |^2= |\Gamma(1+ix)|^2$, and then~\eqref{A-orth} follows again
from~\eqref{orth-S}. Finally, for $m$ even and $n$ odd (or vice versa), \eqref{A-orth} follows trivially
from the fact that $w(x)$ is an even function and $A_m(x) A_n(x)$ is an odd function of $x$.
Note that, due to the identity $\Gamma(ix)/\Gamma(2ix) = 2^{1-2ix}\sqrt{\pi}/\Gamma(\frac12+ix)$, the
function $w$ can be rewritten as
\begin{equation}
w(x)=\left| \frac{\Gamma(a+ix)\Gamma(c+ix)}{\Gamma(\frac12+ix)} \right|^2.  \label{w1}
\end{equation}

Similarly as in~\eqref{phi-n}, the support of the orthogonal polynomials appearing in \eqref{A-orth} is ${\mathbb R}$,
so the spectrum of ${\hat q}$ is ${\mathbb R}$.
Rewriting $\sqrt{w(x)}A_n(x)$ by $\psi_n^{(a,c)}(x)$ (and rescaling the eigenvectors
\eqref{v(x)} by~$\sqrt{w(x)}$), we have the following:

\begin{theorem}\label{theorem1}
In the $su(1,1)_\gamma$ representation ${\cal H}_a$, with $\gamma=(2a-1)(2c-1)$,
the position opera\-tor~${\hat q}$ has formal eigenvectors
\begin{equation*}
v(x)= \sum_{n=0}^\infty \psi_n^{(a,c)}(x)\, |a;n\rangle
\end{equation*}
with
\begin{gather}
 \psi_{2n}^{(a,c)}(x)= \sqrt{w(x)}\frac{(-1)^n S_n(x^2;a,0,c)}{\sqrt{\Gamma(n+a)\Gamma(n+c)\Gamma(n+a+c) n!}},  \label{psi-even}\\
 \psi_{2n+1}^{(a,c)}(x)= \sqrt{w(x)}\frac{(-1)^n   x   S_n(x^2;a,1,c)}{\sqrt{\Gamma(n+a+1)\Gamma(n+c+1)\Gamma(n+a+c) n!}}, \label{psi-odd}
\end{gather}
where $w(x)$ is given in~\eqref{w1}.
The spectrum of ${\hat q}$ is ${\mathbb R}$.
The functions $\psi_n^{(a,c)}(x)$ are orthonormal:
\begin{equation}
\int_{-\infty}^{+\infty} \psi_m^{(a,c)}(x) \psi_n^{(a,c)}(x) dx = \delta_{mn}.
\label{orth-psi}
\end{equation}
The formal eigenvectors satisfy the Dirac delta orthonormality:
\begin{gather*}
 \sum_{n=0}^\infty \psi_n^{(a,c)}(x) \psi_n^{(a,c)}(x') = \delta(x-x'),\qquad
  \int_{-\infty}^{+\infty} v(x) v(x') dx = \delta(x-x').
\end{gather*}
\end{theorem}

The last assertion of the theorem follows from the uniqueness of the weight function
for continuous dual Hahn polynomials~\cite{Ismail1989}
(or, otherwise said, the Hamburger moment problem for continuous dual Hahn polynomials is determinate).

This result also implies that the functions $\psi_n^{(a,c)}(x)$ can be interpreted as position wave functions
for the $\su(1,1)_\gamma$ oscillator model, when the oscillator is in the stationary state $|a,n\rangle$
with energy $n+a$.
This interpretation makes physically sense provided $|\psi_n^{(a,c)}(x)|^2\leq 1$.
Investigating plots of the functions $\psi_n^{(a,c)}(x)$ indicate
that the maximum value for $|\psi_n^{(a,c)}(x)|^2$
is obtained for $|\psi_0^{(a,c)}(0)|^2$. It is easy to see from~\eqref{w1} and~\eqref{psi-even} that
\begin{equation*}
\psi_0^{(a,c)}(0) = \sqrt{\frac{\Gamma(a)\Gamma(c)}{\pi\Gamma(a+c)}} = \sqrt{\frac{B(a,c)}{\pi}},
\end{equation*}
in terms of the Beta function.
So the parameters should satisfy $B(a,c)\leq \pi$.
As can be seen from the Beta integral formula~\cite[(1.1.21)]{Andrews},
this is certainly satisf\/ied when both
$a\geq \frac12$ and $c\geq \frac12$.

\begin{figure}[t!]
\centering
\includegraphics[scale=0.64]{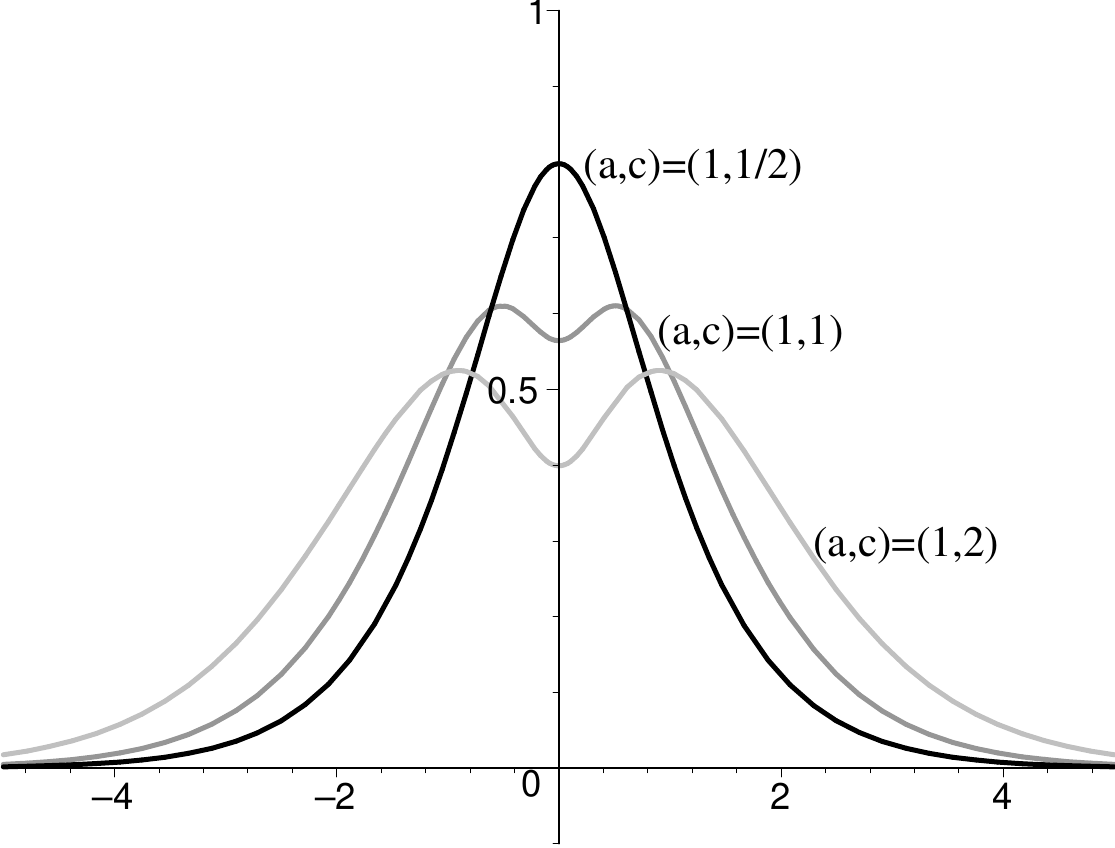} \qquad \includegraphics[scale=0.64]{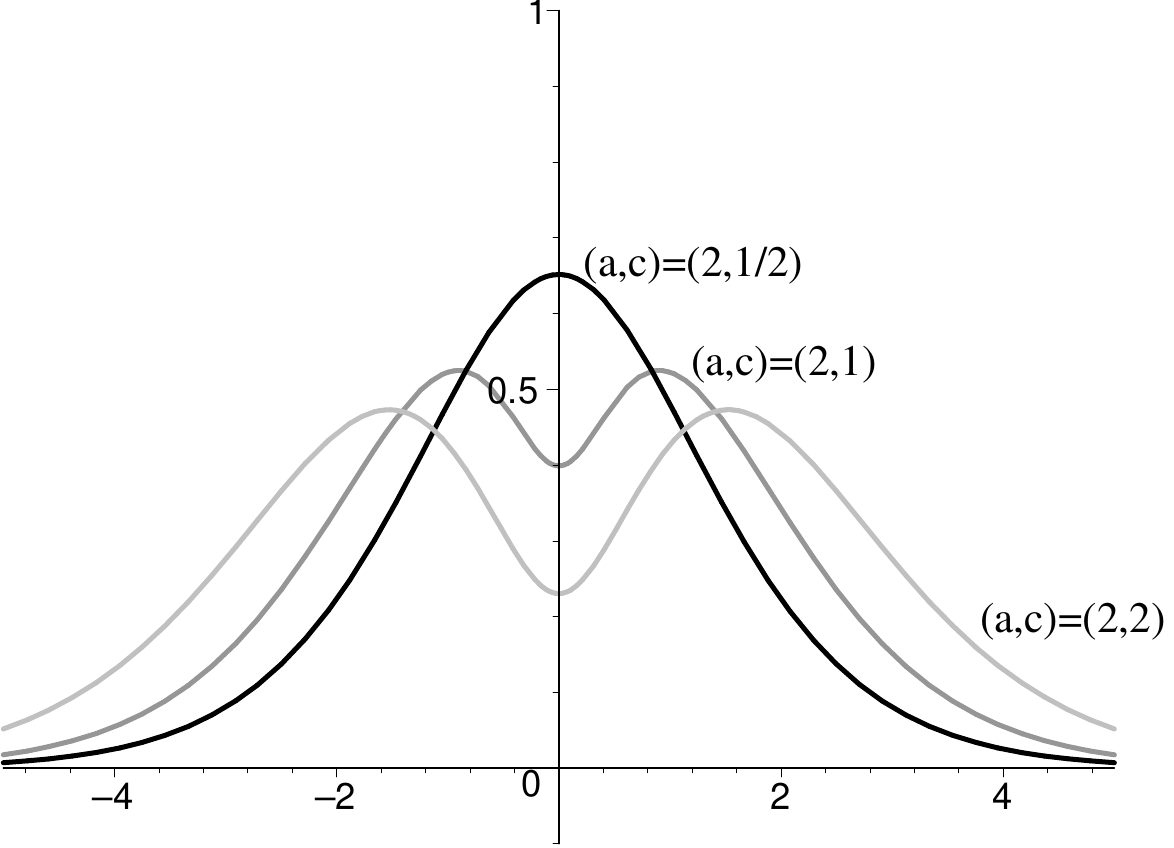}

\caption{Plots of the wave functions $\psi^{(a,c)}_n(x)$ for $n=0$ (ground level), for $a=1$ (left f\/igure) and for
$a=2$ (right f\/igure).
The values of $c$ varies: $c=1/2$ (black), $c=1$ (dark gray) and $c=2$ (light gray).}
\label{fig1}
\end{figure}

\begin{figure}[t!]
\centering
\includegraphics[scale=0.64]{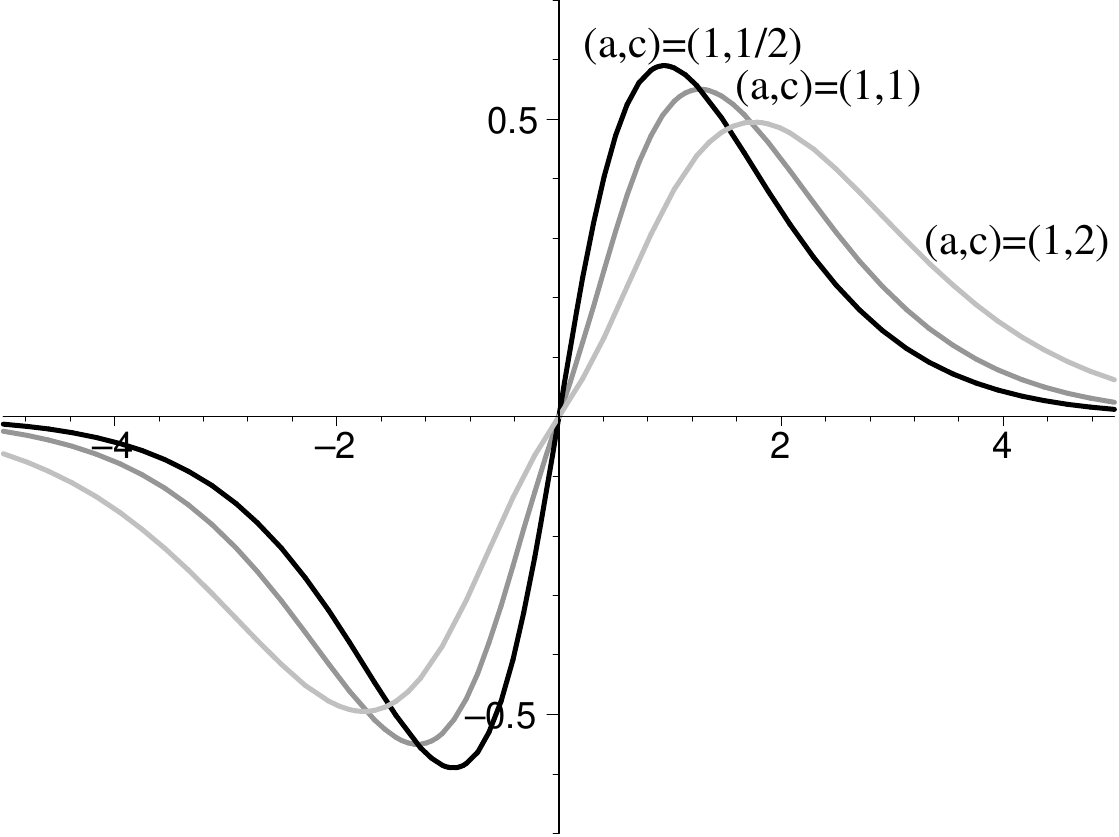} \qquad \includegraphics[scale=0.64]{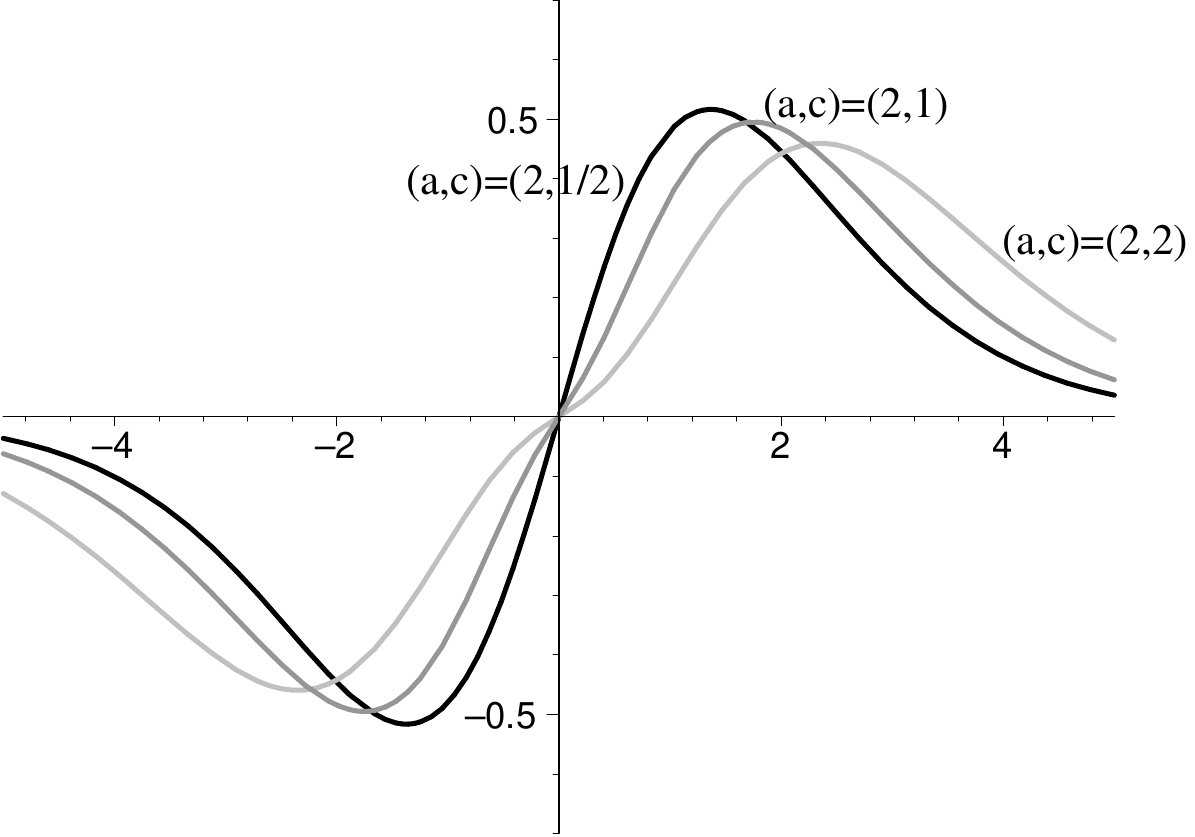}

\caption{Plots of the wave functions $\psi^{(a,c)}_n(x)$ for $n=1$, for $a=1$ (left f\/igure) and for
$a=2$ (right f\/igure).
The values of $c$ are again $c=1/2$ (black), $c=1$ (dark gray) and $c=2$ (light gray).}
\label{fig2}
\end{figure}

It is interesting to study some plots of the wave functions $\psi_n^{(a,c)}(x)$.
First of all, note that this function is symmetric in $a$ and $c$, so one can keep one parameter $a$ f\/ixed and let
the other one $c$ vary.
In Fig.~\ref{fig1} we plot the wave functions for $n=0$ and in Fig.~\ref{fig2} for $n=1$, both for $a=1$ and for $a=2$, for some
values of $c$.
For $c=1/2$ (undeformed algebra), one observes for the ground state wave function a shape similar to the Gaussian function,
but with increasing variance as $a$ increases.
When $c$ increases, the bell shapes are deformed in the middle, and the position probability decreases around the origin.
The deformations for $n>0$ follow a similar pattern.

In a similar way, the spectrum and eigenvalues of the momentum operator ${\hat p}$ can be determined.
Denoting the formal eigenvector of ${\hat p}$, for the eigenvalue $p$, by
\begin{equation*}
\bar v(p) = \sum_{n=0}^\infty {\bar A}_n(p) \, |a,n\rangle,
\end{equation*}
the equation ${\hat p}  \bar v(p) = p  \bar v(p)$ leads, using~\eqref{qpH}, to a set of
recurrence relations similar to~\mbox{\eqref{A1}--\eqref{A2}}.
The solution to these equations is the same as before, up to a multiple of $i$.
So one f\/inds that the formal eigenvectors of ${\hat p}$ are given by
\begin{equation*}
\bar v(p) = \sum_{n=0}^\infty i^n \psi_n^{(a,c)}(p) \, |a,n\rangle,
\end{equation*}
where the functions $\psi_n^{(a,c)}$ are the same as in Theorem~\ref{theorem1}. The spectrum of ${\hat p}$ is ${\mathbb R}$.

\section{Limits and special cases}\label{section5}

For $c=1/2$ we have that $\gamma=0$ and hence the algebra $\su(1,1)_\gamma$ is undeformed.
So for $c=1/2$, the position wave functions $\psi_n^{(a,c)}(x)$ should reduce to the functions $\phi_n^{(a)}(x)$
of Section~\ref{section2}.
This can indeed be verif\/ied explicitly, although it follows already from the fact that putting $c=1/2$ in the
weight function~\eqref{w1} reduces to the corresponding weight function of Meixner--Pollaczek polynomials.

Explicitly, let us f\/irst consider the case with an even index, i.e.\ expression~\eqref{psi-even} for $\psi_{2n}^{(a,1/2)}(x)$.
For the Gamma functions in the denominator of~\eqref{psi-even}, of the form
$\Gamma(n+a)\Gamma(n+a+\frac12) \Gamma(n+\frac12)\Gamma(n+1)$, one can use the famous
multiplication formula
\[
\Gamma(z)\Gamma\left(z+\frac12\right) = 2^{1-2z} \sqrt{\pi} \Gamma(2z).
\]
For the function $S_n(x^2;a,0,1/2)$ in the numerator of~\eqref{psi-even}, one uses
\begin{equation}
{}_3F_2 \left( \myatop{-n,a+ix,a-ix}{a,a+1/2} ; 1 \right) =
{}_2F_1 \left( \myatop{-2n,a+ix}{2a} ; 2 \right),
\label{FF1}
\end{equation}
and this last expression ${}_2F_1$ yields the Meixner--Pollaczek polynomial in~\eqref{defP}.
So one f\/inds indeed:
\begin{equation*}
\psi_{2n}^{(a,1/2)}(x) = \phi_{2n}^{(a)}(x).
\end{equation*}
Note that the identity~\eqref{FF1} is obtained by f\/irst applying a Thomae--Weber--Erdelyi transformation
on the ${}_3F_2$~\cite[Appendix]{Rao} (sometimes called a Shephard transformation~\cite[Corolla\-ry~3.3.4]{Andrews}):
\[
{}_3F_2 \left( \myatop{-n,a+ix,a-ix}{a,a+1/2} ; 1 \right) = \frac{(\frac12)_n}{(a+\frac12)n}
\,{}_3F_2 \left( \myatop{-n,ix,-ix}{a,1/2} ; 1 \right),
\]
and then one can turn this last ${}_3F_2$ into a ${}_2F_1$ by~\cite[equation~(39)]{JSV2011b}.

For an odd index, one uses expression~\eqref{psi-odd} for $\psi_{2n+1}^{(a,1/2)}(x)$.
The computation is similar, the main dif\/ference coming from the appearance of
$S_n(x^2;a,1,1/2)$ in the numerator of~\eqref{psi-odd}.
Here, one can use
\begin{equation}
{}_3F_2 \left( \myatop{-n,a+ix,a-ix}{a+1,a+1/2} ; 1 \right) = \frac{ia}{x}
\,{} _2F_1 \left( \myatop{-2n-1,a+ix}{2a} ; 2 \right).
\label{FF2}
\end{equation}
The last identity~\eqref{FF2} is again obtained by f\/irst applying a Thomae--Weber--Erdelyi transformation
on the ${}_3F_2$~\cite[Appendix]{Rao}:
\[
{}_3F_2 \left( \myatop{-n,a+ix,a-ix}{a+1,a+1/2} ; 1 \right) = \frac{(\frac32)_n}{(a+\frac12)n}
\,{}_3F_2 \left( \myatop{-n,1+ix,1-ix}{a+1,3/2} ; 1 \right),
\]
and then using~\cite[equation~(40)]{JSV2011b}.

A second interesting case is the limit $c\rightarrow +\infty$.
Note from the action~\eqref{act-J+}--\eqref{act-J-} that the operators
\begin{equation*}
b^+ = \lim_{c\rightarrow\infty} \frac{J_+}{\sqrt{2c}},\qquad
b^- = \lim_{c\rightarrow\infty} \frac{J_-}{\sqrt{2c}},
\end{equation*}
have the same action on ${\cal H}_a$ as the paraboson oscillator creation and
annihilation operators~\cite[equation~(A7)]{JSV2011}.
Under this same limit, the position and momentum operators become
\begin{equation}
\hat Q = \lim_{c\rightarrow\infty} \frac{J_+ +J_-}{2\sqrt{c}},\qquad
\hat P = \lim_{c\rightarrow\infty} i\frac{J_+ -J_-}{2\sqrt{c}}
\label{QP}
\end{equation}
and these operators satisfy, from~\eqref{J+J-} and~\eqref{qpH},
\begin{gather*}
[\hat Q, \hat P] = \lim_{c\rightarrow\infty} \left(\frac{i}{c}J_0+\frac{i\gamma}{2c}R\right) =
\lim_{c\rightarrow\infty} \left(\frac{i}{c}\left(\hat H +c-\frac12\right)+\frac{i(2a-1)(2c-1)}{2c}R\right) \\
\hphantom{[\hat Q, \hat P]}{} = i+(2a-1)iR,
\end{gather*}
which is the common commutator in the paraboson case~\cite{Ohnuki,Ply,Vinet1}.
Hence one can also expect the wave functions $\psi_n^{(a,c)}(x)$ to tend to the
paraboson wave functions $\Psi_n^{(a)}(x)$ \cite[equation~(A.11)]{JSV2011} when
$c$ tends to $\infty$.
In order to consider this limit, note that the position operator has been divided by $\sqrt{c}$ in~\eqref{QP},
so for its eigenvalue we should introduce a new variable by putting $x=\sqrt{c}  \xi$.
Then according to~\eqref{orth-psi}, we need to compute the limit
\begin{equation*}
\lim_{c\rightarrow +\infty} c^{1/4} \psi_n^{(a,c)} \big(\sqrt{c}  \xi\big).
\end{equation*}

Let us consider the even case $\psi_{2n}^{(a,c)}$ (the odd case is similar).
For the polynomial part, \eqref{psi-even}~and~\eqref{defS} lead to the following limit:
\begin{equation*}
\lim_{c\rightarrow \infty} \,{}_3F_2 \left( \myatop{-n,a+i\sqrt{c} \xi,a-i\sqrt{c} \xi}{a,a+c} ; 1 \right) =
\,{}_1F_1 \left( \myatop{-n}{a} ; \xi^2 \right)
= \frac{n!}{(a)_n} L_n^{(a-1)} \big(\xi^2\big),
\end{equation*}
where $L_n^{(a-1)}$ is a Laguerre polynomial~\cite{Koekoek}.
Taking care of the other factors, one f\/inds
\begin{equation*}
\lim_{c\rightarrow \infty} c^{1/4} \psi_{2n}^{(a,c)} \big(\sqrt{c}  \xi\big)
=  (-1)^n  \sqrt {\frac{n!}{\Gamma( n + a) } }
	|\xi|^{a-1/2}   e^{-\xi^2/2} L_n^{(a-1)}\big(\xi^2\big).
\end{equation*}
So one f\/inds indeed
\begin{equation*}
\lim_{c\rightarrow +\infty} c^{1/4} \psi_n^{(a,c)} \big(\sqrt{c}  \xi\big) =
\Psi_n^{(a)}(\xi),
\end{equation*}
in terms of the paraboson wave functions~\cite[equation~(A.11)]{JSV2011}.

Note that for $a=1/2$ the paraboson wave functions reduce to the canonical oscillator wave functions,
as the corresponding Laguerre polynomials become Hermite polynomials~\cite[Appendix]{JSV2011}.
So therefore, one can say that under the limit $(a,c)\rightarrow (1/2,+\infty)$ the new oscillator models
introduced in this paper reduce to the canonical quantum oscillator.

\section[Differential-reflection operator realization of $\su(1,1)_\gamma$]{Dif\/ferential-ref\/lection operator realization of $\boldsymbol{\su(1,1)_\gamma}$}\label{section6}

Just as $\su(1,1)$ has a dif\/ferential operator realization, and a realization of the
Hilbert spa\-ce~${\cal H}_a$~\cite{Klimyk2006}, the analogue can be constructed for $\su(1,1)_\gamma$ as well.
The basis vectors can be realized as monomials in a variable $z$:
\begin{equation}
|a,2n\rangle = \sqrt{\frac{(a)_n(a+c)_n}{(c)_n n!}}\ z^{2n}, \qquad
|a,2n+1\rangle = \sqrt{\frac{(a)_{n+1}(a+c)_n}{(c)_{n+1} n!}}\ z^{2n+1}.
\label{z}
\end{equation}
Clearly, the `abstract' ref\/lection operator $R$ is realized by the concrete ref\/lection operator $\breve{R}$ with action
\begin{equation*}
\breve{R}  f(z) = f(-z).
\end{equation*}
Then it is a simple exercise to verify that the following dif\/ferential-ref\/lection operators satisfy
\eqref{act-J0}--\eqref{act-J-} when acting on the monomials~\eqref{z}
\begin{gather*}
  J_0  = z \frac{d}{dz} +a+c-\frac12, \qquad
  J_- = \frac{d}{dz} + \left(c-\frac12\right) \frac{1-\breve{R}}{z}, \\ 
 J_+ = z^2 \frac{d}{dz} + 2a z + \left(c-\frac12\right)z (1-\breve{R}).
\end{gather*}
In order to characterize the Hilbert space~${\cal H}_a$ as a space of analytic functions,
a description of the scalar product should be found in such a way that the monomials~\eqref{z}
are orthonormal.
So far, we have not been able to construct this.
Let us nevertheless mention that the position (and momentum) wave functions have a more explicit
expression in this realization.
Indeed,
\begin{equation}
v(x) = \sum_n \psi_n^{(a,c)}(x) \, |a,n\rangle =
\sum_n \psi_{2n}^{(a,c)}(x) \, |a,2n\rangle + \sum_n \psi_{2n+1}^{(a,c)}(x) \,|a,2n+1\rangle.
\label{vxz}
\end{equation}
For the f\/irst sum, one f\/inds
\begin{gather*}
\sum_n \psi_{2n}^{(a,c)}(x) \, |a,2n\rangle  =
\sqrt{\frac{w(x)}{\Gamma(a)\Gamma(c)\Gamma(a+c)}} \sum_n \frac{S_n(x^2;a,0,c)}{(c)_n n!} \big({-}z^2\big)^n \nonumber\\
 \hphantom{\sum_n \psi_{2n}^{(a,c)}(x) \, |a,2n\rangle}{} = \sqrt{\frac{w(x)}{\Gamma(a)\Gamma(c)\Gamma(a+c)}} \big(1+z^2\big)^{-a+ix}
\,{} _2F_1 \left( \myatop{ix,c+ix}{c} ; -z^2 \right),
\end{gather*}
by using~\cite[(9.3.14)]{Koekoek}.
In a similar way, the second sum of~\eqref{vxz} equals
\begin{equation*}
\sqrt{\frac{w(x)}{\Gamma(a)\Gamma(c)\Gamma(a+c)}} \frac{xz}{c} \big(1+z^2\big)^{-a+ix}
\,{}_2F_1 \left( \myatop{1+ix,c+ix}{1+c} ; -z^2 \right).
\end{equation*}

\section[A further deformation of $\su(1,1)_\gamma$]{A further deformation of $\boldsymbol{\su(1,1)_\gamma}$}\label{section7}

The model in Section~\ref{section4} has two parameters $a$ and $c$: one coming from the representation label~$a$, and
one coming from the deformation parameter $\gamma=(2a-1)(2c-1)$.
The position operator eigenfunctions are in terms of continuous dual Hahn polynomials $S_n(x^2;a,0,c)$ or
$S_n(x^2;a,1,c)$, due to the relations~\eqref{S10}--\eqref{S20}.
However, continuous dual Hahn polynomials $S_n(x^2;a,b,c)$ have in general three parameters~$a$,~$b$ and~$c$, and
furthermore the relations~\eqref{S1}--\eqref{S2} preceding~\mbox{\eqref{S10}--\eqref{S20}} are indeed in terms
of three parameters.
So one may wonder whether a third deformation parameter $b$ could be introduced in the algebraic relation~\eqref{J+J-}.
This is in fact the case; however, we shall see that it leads to `unphysical' wave functions.

Assume that we def\/ine a new deformed algebra as a unital algebra with elements
$J_0$, $J_+$, $J_-$ and $R$ subject to the relations of Def\/inition~\ref{definition1}, but with~\eqref{J+J-} replaced by:
\begin{equation*}
[J_+,J_-] = -2J_0 -\gamma R -4b J_0 R,
\end{equation*}
where as before $\gamma=(2a-1)(2c-1)$.
The actions of Proposition~\ref{proposition2} on a representation space~${\cal H}_a$ can then be generalized to{\samepage
\begin{gather*}
  J_0 |a,n\rangle = \left(n+a+b+c-\frac12\right) |a,n\rangle,\\ 
  J_+ |a,n\rangle =
  \begin{cases}
 \sqrt{(n+2a+2b)(n+2b+2c)}\,|a,n+1\rangle, & \text{if $n$ is even,}\\
 \sqrt{(n+1)(n+2a+2c-1)}\,|a,n+1\rangle, & \text{if $n$ is odd,}
 \end{cases} \\ 
  J_- |a,n\rangle =
  \begin{cases}
 \sqrt{n(n +2a+2c-2)}\,|a,n-1\rangle, & \text{if $n$ is even,}\\
 \sqrt{(n+2a+2b-1)(n+2b+2c-1)}\,|a,n-1\rangle, & \text{if $n$ is odd,}
 \end{cases} \label{act-J-b}
\end{gather*}
and the action of $R$ is unchanged. We shall also assume that $a,c>0$ and $b\geq 0$.}

In this case, the expressions \eqref{qpH} can still be used, and thus a formal eigenvector of $\hat q$ for the eigenvalue~$x$,
of the form~\eqref{v(x)}, leads to:
\begin{gather*}
  x  A_{2n}(x) = \sqrt{(n+a+b)(n+b+c)} A_{2n+1}(x) + \sqrt{n(n+a+c-1)} A_{2n-1}(x),\\ 
  x  A_{2n+1}(x) = \sqrt{(n+1)(n+a+c)} A_{2n+2}(x) + \sqrt{(n+a+b)(n+b+c)} A_{2n}(x), 
\end{gather*}
reminiscent of the more general equations~\eqref{S1}--\eqref{S2}.
These equations can indeed be solved by taking
\begin{gather*}
  A_{2n}(x)= \frac{(-1)^n S_n(x^2-b^2;a,b,c)}{\sqrt{\Gamma(n+a+b)\Gamma(n+b+c)\Gamma(n+a+c) n!}}, \\ 
  A_{2n+1}(x)= \frac{(-1)^n \; x \; S_n(x^2-b^2;a,b+1,c)}{\sqrt{\Gamma(n+a+b+1)\Gamma(n+b+c+1)\Gamma(n+a+c) n!}}. 
\end{gather*}
Although this is a formal solution, the appearance of $x^2-b^2$ as an argument of the continuous dual Hahn polynomials
spoils the orthogonality relation~\eqref{orth-S}, and we would not obtain~${\mathbb R}$ as spectrum of the position operator
(but rather the values for which $|x|>|b|$, plus possibly some discrete points according to the
orthogonality~\cite[(9.3.3)]{Koekoek}, depending on the values of~$a$,~$b$,~$c$).
Because of the unphysical nature of the corresponding eigenfunctions, we will not consider this further.

\section{Conclusion}\label{section8}

The one-dimensional quantum harmonic oscillator is a central problem and model in quantum mechanics.
The simplest and standard model, the non-relativistic quantum harmonic oscillator in the canonical approach,
has an attractive solution for its wave functions of stationary states in terms of Hermite polynomials.
The dynamical algebra of this standard oscillator (or ``Hermite oscillator'') is the usual Heisenberg algebra.

This standard model can be extended both for continuous and discrete measures (for the wave functions), and
in both cases some elegant models with analytic solutions for the wave functions exist.

In the continuous case, there are two well-known ways of extending the Hermite oscillator: these are the
Meixner--Pollaczek oscillator~\cite{Klimyk2006} and the paraboson oscillator~\cite{Ohnuki}.
Both extensions have the same equidistant energy spectrum, where the ground state energy is some positive value $a$
instead of $1/2$ in the case of the Hermite oscillator.
The dynamical algebra is quite dif\/ferent though.
For the Meixner--Pollaczek oscillator, the Hamiltonian together with the position and momentum operator form
a basis of the Lie algebra $\su(1,1)$. The ground state energy $a$ corresponds to the representation label (lowest
weight) of a positive discrete series representation.
For the paraboson oscillator, the position and momentum operators are considered as odd ge\-ne\-ra\-tors of
the Lie superalgebra $\osp(1|2)$. The ground state energy $a$ is again a representation label for a unitary
$\osp(1|2)$ representation. Note that in this case the wave functions are given in terms of (generalized)
Laguerre polynomials.

By constructing models for the quantum harmonic oscillator based upon a deformation of~$\su(1,1)$,
we have been able to unify the previous well-known extensions in the continuous case.
The algebra $\su(1,1)_\gamma$ is an extension of $\su(1,1)$ by a parity operator~$R$, and involves
a deformation parameter $\gamma$. The proposed models for the oscillator involve two parameters $a$ and $c$:
$a$ is again a representation label, and $c$ is a deformation label related to $\gamma$ by $\gamma=(2a-1)(2c-1)$.
The energy spectrum is again equidistant, with ground state energy equal to $a$.
Our main result is that the stationary wave functions of such models have elegant closed form expressions
in terms of continuous dual Hahn polynomials.
Some properties of these wave functions have been described in Section~\ref{section4}.

Clearly, the dynamical algebra for these new models is $\su(1,1)_\gamma$.
For $c=1/2$, $\su(1,1)_\gamma$ becomes $\su(1,1)$ and also the wave functions for the new models reduce to those
of the Meixner--Pollaczek oscillator.
For $c\rightarrow \infty$, the algebra reduces to the paraboson algebra, and the wave functions become the
known paraboson oscillator wave functions.

As far as potential applications to physical models are concerned, we can of\/fer the argument that our proposed model
has two parameters $a$ and $c$ (one more that the Meixner--Pollaczek oscillator and the paraboson oscillator, and
two more than the Hermite oscillator).
These parameters should be greater than or equal to $1/2$ in order to deal with physically acceptable wave functions.
Having two parameters available in a mathematical model for the quantum oscillator, opens the way
to more f\/lexibility in applications.
For possible applications (of the deformed algebra and its representations)
in quantum f\/ield theory, see the ideas presented in~\cite{Horvathy}.

\subsection*{Acknowledgments}
E.I.~Jafarov was supported by a postdoc fellowship from the Azerbaijan National Academy of Sciences.
N.I.~Stoilova was supported by project P6/02 of the Interuniversity Attraction Poles Programme (Belgian State --
Belgian Science Policy).

\pdfbookmark[1]{References}{ref}
\LastPageEnding

\end{document}